# Comment on "Bouncing ball with finite restitution: Chattering, locking, and chaos"

Nicholas B. Tufillaro*
Center for Nonlinear Studies and Theoretical Division,
MS-B258, Los Alamos National Laboratories
Los Alamos, New Mexico 87545 USA

(28 February 1994)

Some statements by Luck and Mehta [Phys. Rev. E **48**, 3988 (1993)] drawn from their analysis of the "exact" one-dimensional model of a bouncing ball system are either misleading or incorrect. In particular, in agreement with previous theoretical and experimental studies, the bouncing ball system does exhibit chaotic orbits for a wide range of experimentally accessible parameters. The "sticking solutions" with long transients analyzed by Luck and Mehta are also observed and usually easily distinguished from the chaotic orbits.

46.10.+z, 03.20.+i, 05.45.+b

Several researchers have studied one-dimensional models of the bouncing ball system which include the coefficient of restitution ($0 \leq \alpha \leq 1$), and many have also noted the existence of the large class of eventually periodic orbits known as "sticking solutions" [1]. More references can be found in Ref. [2]. All these models—which are equivalent to the dynamical equations described by Luck and Mehta—have been termed the "exact" one-dimensional model of the bouncing ball system [2]. The phrase "one-dimensional" refers to the number of degrees of freedom the ball moves in and not to the dimension of the phase space model.

To fix a notation which allows an easier comparison with experiments, recall that the dynamics of the bouncing ball system can be found by solving the (implicit) nonlinear coupled algebraic equations known as the *phase map*,

$$A[\sin(\theta_k) + 1] + v_k \left[\frac{1}{\omega}(\theta_{k+1} - \theta_k)\right]$$
$$-\frac{1}{2}g\left[\frac{1}{\omega}(\theta_{k+1} - \theta_k)\right]^2 - A[\sin(\theta_{k+1}) + 1] = 0 \quad (1)$$

and the *velocity map*,

$$(1+\alpha)\omega A \cos(\theta_{k+1}) - \alpha\left\{v_k - g\left[\frac{1}{\omega}(\theta_{k+1}-\theta_k)\right]\right\} = v_{k+1} \quad (2)$$

where $\theta_k = \omega t + \theta_0$ and $v_k$ are the phase and velocity of the k-th impact between the ball and oscillating table, $A$ and $\omega$ are the table's amplitude and angular frequency, $\alpha$ is the coefficient of restitution, and $g$ is the gravitational acceleration. The implicit phase map and explicit velocity map constitute the *exact* model of the bouncing ball system. Earlier experimental studies showed an excellent correspondence between the exact model and the dynamics of an experimental bouncing ball system, all the major bifurcations predicted by the exact model occurred within the experimental system [3]. Observations between the model and experiment agreed to within %2 with no fitted parameters. A public domain program, the *Bouncing Ball Simulation System*, has also been available since 1986 which simulates the exact model [2]. We use this program to obtain the results presented here.

Experiments illustrating chaos in the bouncing ball system usually proceed along the following lines. The amplitude the table driving the ball is slowly increased while monitoring the dynamics of the bouncing ball through an experimental impact map, which is similar to a next return map [2]. In essence, an experimental bifurcation diagram is created. The coefficient of restitution can be changed from around 0.2 to 0.8 by using different materials for the ball (eg., wood, plastic, steel). Experimentally, it is observed that a chaotic invariant set is seen at the end of the period doubling cascade, but for a further increase in the driving amplitude, *the strange attractor is destroyed by a crisis*. The dynamics of the ball after this crisis can result in motion which can quickly approach a periodic sticking solution (generally speaking, for smaller values of $\alpha$), or can exhibit long transients—sometimes called 'transient chaos' [5]—following the "shadow of the strange attractor" (generally speaking, for larger values of $\alpha$). It is the dynamics of this transient chaos when $\alpha$ is close to one that Luck and Mehta analyze [6].

Direct simulation of the "exact" model exhibits a similar behavior. Figure 1 presents a bifurcation diagram showing a period doubling route to chaos for $\alpha = 0.5$. Note that this strange attractor is approached in exactly the same way as it would be in an experiment, namely, by slowly scanning the amplitude until the end of the period doubling cascade is reached and a non-periodic orbit is observed. In simulations ($A = 0.012$) the strange attractor is found to be stable for over $10^6$ impacts. Between $A = 0.0121$ and $A = 0.0122$ a crisis occurs which destroys this strange attractor. For $A > 0.0122$ the orbit follows the shadow of the strange attractor for a number of impacts but eventually converges to a sticking solution (typically after $10^2$ to $10^3$ impacts). In both experiments and simulations, the pre-crisis (chaotic) dynamics and post-crisis (eventually periodic) dynamics are usually easy to distinguish because the range of impact phases explored by the ball suddenly widens after the crisis. In the simulation shown in Figure 1, the chaotic dynamics is confined to a phase between $-0.1 < \theta/2\pi < 0.3$ where as the post-crisis dynamics explores almost the entire range of phases available. This feature provides a nice signature to distinguish the pre- and post-crisis dynamics in both experiments and simulations.

This general scenario of period-doubling, chaos, crisis, and sticking solutions (possibly with transient chaos) is not confined to a few selected parameter values but is generally observed for a wide range of $\alpha$. For instance, Fig. 2 shows this same scenario for $\alpha = 0.1$, and Fig. 3 for $\alpha = 0.8$. Figure 3, though, also illustrates that the amplitude range where a strange attractor is observed shrinks as $\alpha$ approaches one, which is perhaps the reason why Luck and Mehta did not notice this scenario, especially if they confined their simulations and analysis to the regime where $\alpha \approx 1$.

We conclude by stating that in our opinion earlier theory and experiments did not take a "rather cavalier" attitude toward models including a finite coefficient of restitution, but that when presenting results of earlier experiments [3,7] the experimenters where fully aware of the coexistence of strange attractors and sticking solutions, and how to experimentally distinguish both types of invariant sets. Further, while we agree that one-dimensional maps are in general a good qualitative first step in modeling dynamical systems which are inherently two-dimensional, we must also disagree with the statement that the completely inelastic ($\alpha = 0$) model is "a good qualitative indicator for the dynamics of the ball with finite restitution up to values close to 1." Rather, our experiments and simulations show significant new behavior in the exact model which is not predicted by the completely inelastic model for moderate inelasticity (say, $\alpha = 0.5$). Finally, we find that a strange attractor is easy to observe in both experiments and simulations for realistic and experimentally accessible parameter values.

In a future communication we will actually use topological methods to "prove" the existence of a chaotic invariant set in the exact one-dimensional model of the bouncing ball system [8].

FIG. 1. Bifurcation diagram for the exact one-dimensional model of the bouncing ball system with damping $\alpha = 0.5$. The table amplitude (A) is measured in centimeters. Diagram generated with the Bouncing Ball Simulation System. For more details about the program see Ref. [2].

FIG. 2. Bifurcation diagram for the exact one-dimensional model of the bouncing ball system with damping $\alpha = 0.1$. Diagram generated with the Bouncing Ball Simulation System [2].

FIG. 3. Bifurcation diagram for the exact one-dimensional model of the bouncing ball system with damping $\alpha = 0.8$. Diagram generated with the Bouncing Ball Simulation System [2].

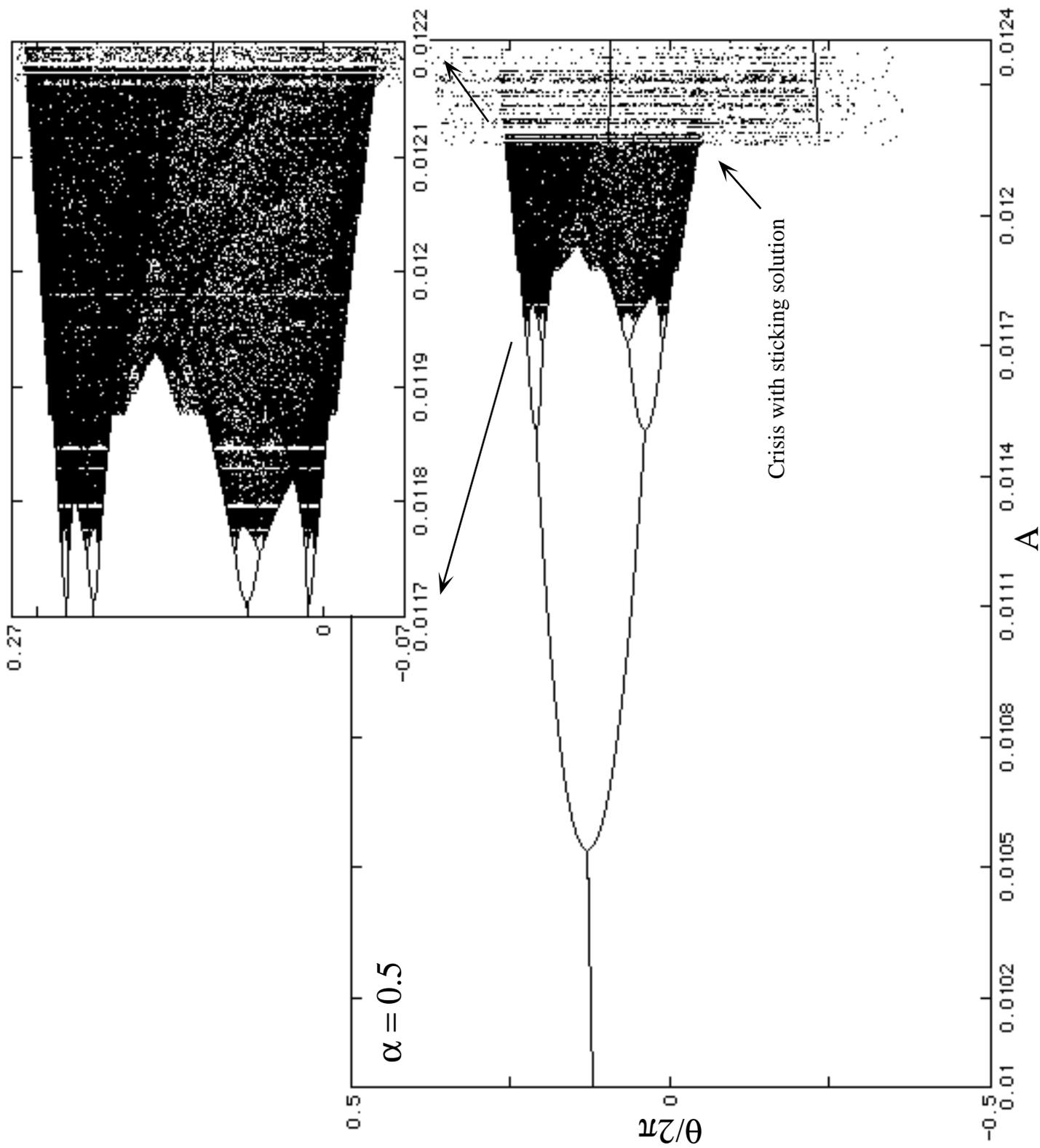

Figure 1. Tufillaro: Comment on ``Bouncing ball with finite restitution: Chattering, locking, and c

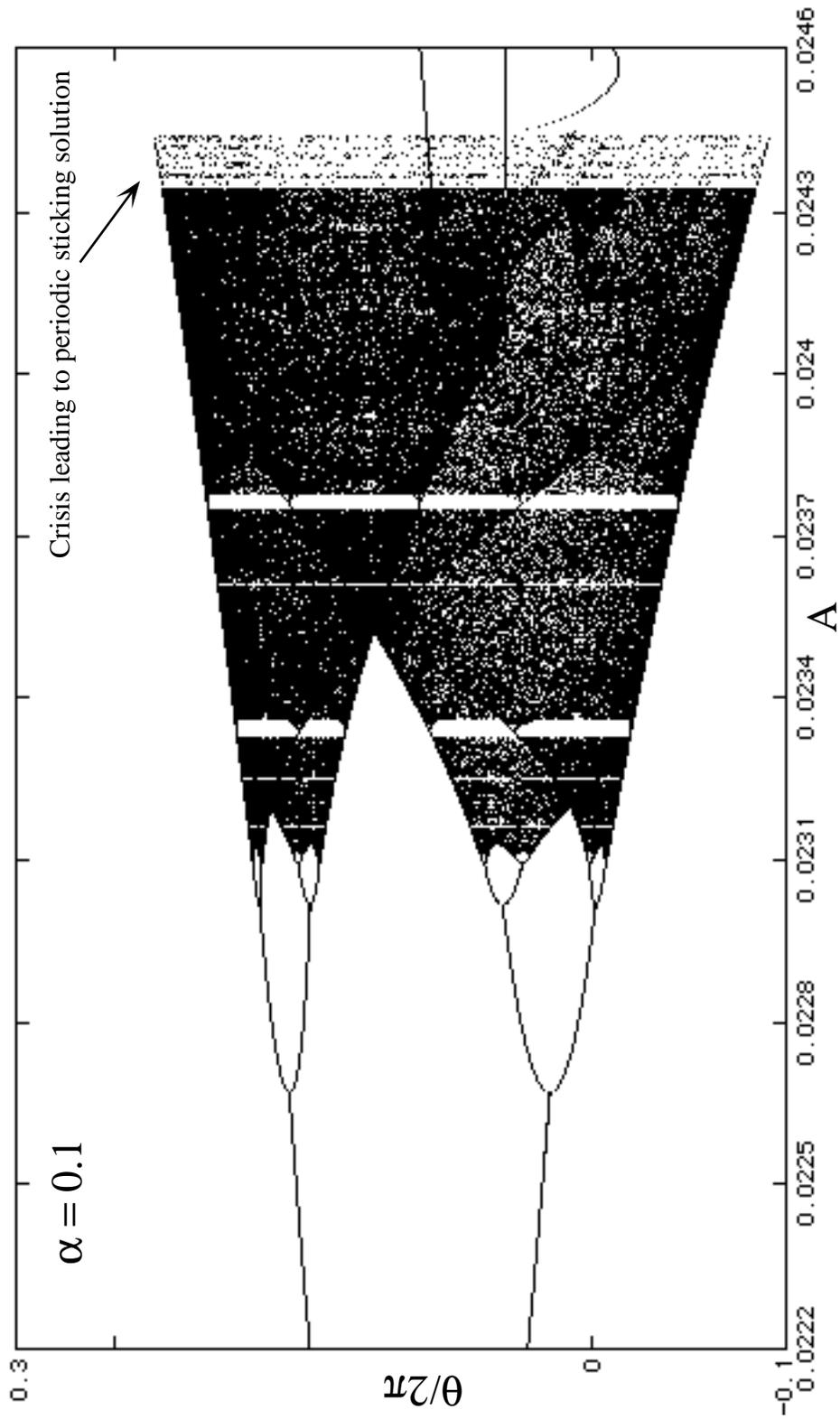

Figure 2. Tufillaro: Comment on ``Bouncing ball with finite restitution: Chattering, locking, and

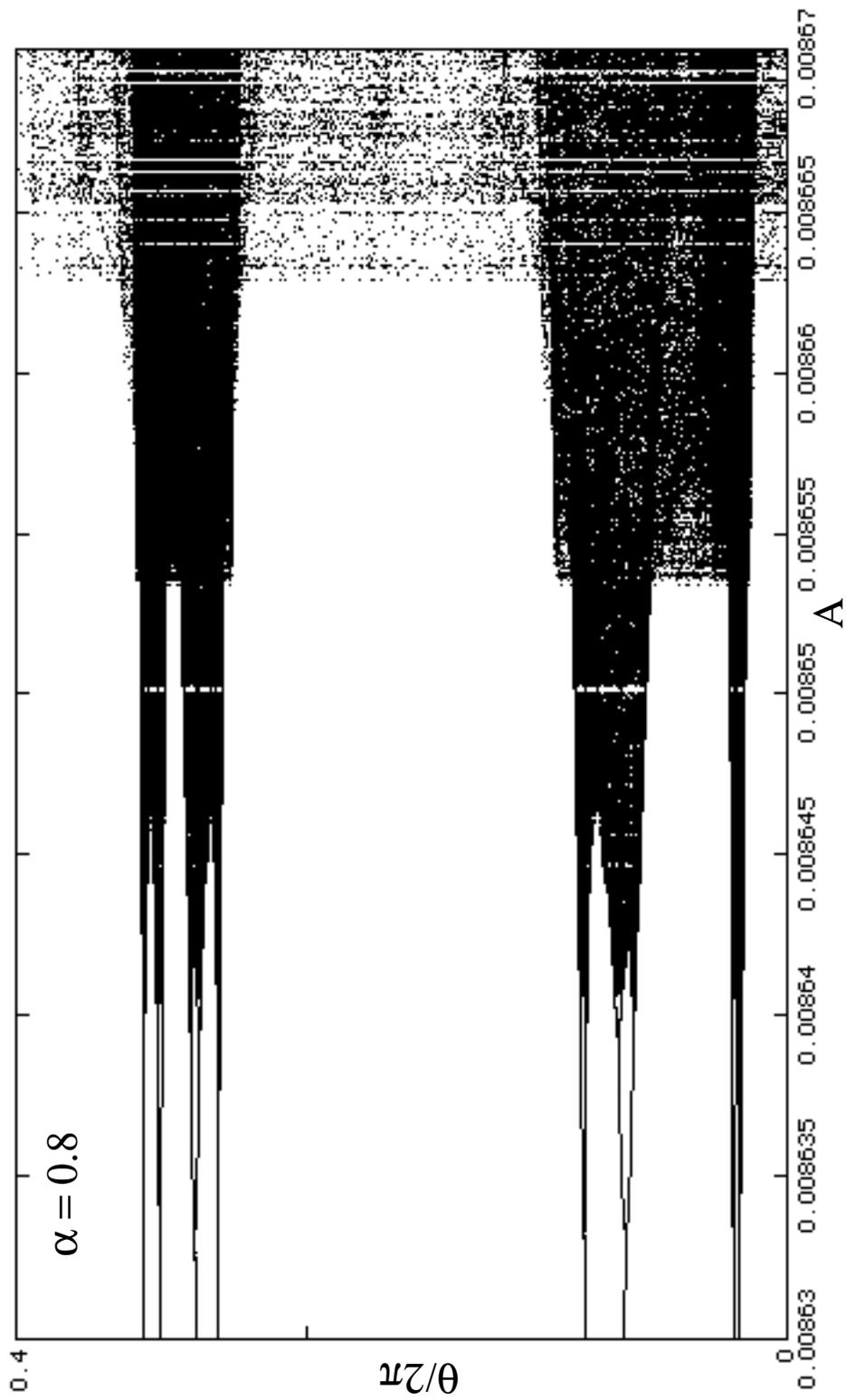

Figure 3. Tufillaro: Comment on ''Bouncing ball with finite restitution: Chattering, locking, and